\documentclass[twocolumn,showpacs,preprintnumbers,amsmath,amssymb]{revtex4}
\usepackage[dvips]{graphicx}

\begin{document}
\title{Simulations of prompt many-body ionization in a frozen Rydberg gas}
\author{F. Robicheaux}
\affiliation{Department of Physics, Purdue University, West Lafayette,
Indiana 47907, USA}
\email{robichf@purdue.edu}
\author{M. M. Goforth}
\author{M. A. Phillips}
\affiliation{Department of Physics, Auburn University, AL
36849-5311}
\date{\today}

\begin{abstract}
The results of a theoretical investigation of prompt many-body ionization
are reported. Our calculations address an experiment
that reported ionization in Rydberg
gases for densities two orders of magnitude
less than expected from ionization between pairs of atoms.
The authors argued that the results were due to the simultaneous
interaction between
many atoms. We performed classical calculations
for many interacting Rydberg atoms with the ions fixed in space
and have found that the many atom
interaction does allow ionization at lower densities than
estimates from two atom interactions. However, we found that the density
fluctuations in a gas play a larger role.
These two effects are an order of magnitude too small to
account for the experimental results suggesting at least
one other mechanism strongly affects ionization.
\end{abstract}


\maketitle
\section{Introduction}

Most low density gases (e.g. $10^{10}$~cm$^{-3}$)
consist of nearly independent atoms or molecules that interact
through random binary collisions. Gases consisting of Rydberg atoms
can violate this picture strongly. Even though the sizes of the
atoms are small compared to the spacing between the atoms,
the large dipole moments that can be formed allow for a large
interaction between atoms. There have been many recent experiments
where the atoms have been separated by 10's of $\mu$m, but still
showed strong interaction. The strong interaction between the atoms
and the controllability inherent in exciting specific
states has led to the possibility of using Rydberg gases as
examples of many-body systems.

A recent experiment\cite{THS} found that ionization in a frozen
Rydberg gas occured at much lower densities than expected from
calculations of pair-wise interactions.\cite{FR1} In this calculation,
it was found that 90\%
of the trajectories led to ionization when a pair of atoms were
at a separation of $\sim 2.1\times 2 n^2$~a$_0$. In Ref.~\cite{THS},
they used
\begin{equation}\label{eqDen}
\rho =\frac{1}{(4\pi /3) (4 n^2\; a_0)^3} =\frac{3}{256\pi n^6\; a_0^3}
\end{equation}
as the reference density; this density corresponds to one atom
within a sphere of radius $4 n^2a_0$. For a Rb 45d state, $n\simeq 43.7$ and the
reference density is $\sim 4\times 10^{12}$~cm$^{-3}$. In their
experiment, they excite the atoms to the 45d state, wait 10~ns,
and then ramp an electric field to measure the ions and atoms.
The details of the measurement means that they measure the
ionization approximately ``100~ns after laser excitation of the
frozen Rydberg gas." The moniker ``frozen Rydberg gas" is applicable
because the Rb atoms have a temperature of 300~$\mu$K giving an
RMS speed of $\sim 0.3$~m/s. Thus, during 100~ns, the atoms move
$\sim 30$~nm which is much smaller than the size of the atoms,
$\sim 200$~nm, {\it and} is much, much smaller than the spacing
between atoms, $\sim 1/\rho_{exp}^{1/3}\sim 3000$~nm. Because the atoms
travel such a small distance, they don't ionize through
collisional processes.

At a density of $5\times 10^{10}$~cm$^{-3}$, they measured substantial
ionization at early times. The surprisingly large amount of ionization
was attributed to many-body interactions since the
ionization occurs at densities roughly 2
orders of magnitude less than the base density needed for ionization
between pairs of atoms, Eq.~(\ref{eqDen}). There have been several
other experimental studies of ionization in a Rydberg gas or
the conversion of the Rydberg gas to a plasma (e.g. see
Refs.~\cite{RTN,RH1,PPR,LNR,LTG,ARW,MRK,ZFZ,VHS,SOM}).
However, these studies are fundamentally
different from Ref.~\cite{THS}
in that the time scale of the ionization is much longer
and the atoms move a substantial fraction of their spacing.
There was a quantum calculation of the autoionization
from pairs of Rydberg atoms\cite{ADG}, but the authors
found that this quantum effect was negligible for the
situation of Ref.~\cite{THS}.

In order to test the idea of many-body
ionization, we performed classical trajectory
Monte Carlo calculations of many interacting Rydberg atoms. Since
the atoms are in highly excited states and the physics involves
substantial averaging, we expect that classical calculations will
provide a good approximation to the actual quantum physics. The
advantage of the classical calculation is that we can include
all of the electron-electron interactions without approximation.
Thus, the ionization process will be properly represented
even if it requires the interaction between
many widely spaced atoms.

There is an important difference between classical and quantum
calculations of ionization for a pair of Rydberg atoms. In the
quantum calculation, there is a nonzero matrix element from
the dipole-dipole interaction which can cause one electron
to go to a more deeply bound state and the other electron to
be ionized. Since the ionized electron is in the continuum,
this is the analog of autoionization.\cite{ADG} Thus, for
large separations, $R$, the quantum decay rate decreases like $1/R^6$
since the matrix element is proportional to $1/R^3$. In contrast,
the classical ionization probability becomes 0 outside of
a separation not much larger than $3.5\times$ the atom
size. The classical ionization probability drops exactly to
0 because as one electron gains energy (thus decreasing
its Rydberg frequency) the other electron loses energy
(thus increasing its Rydberg frequency). This is analogous
to driving a pendulum exactly on resonance for small
angle oscillations (if the coupling is weak, the oscillator
gains energy until the oscillation frequency changes enough
to put it out of phase with the drive). The energy of each
atom oscillates around the average energy with a spread
that decreases as the separation of atoms increase.
This difference in ionization at large separation
is not important for the calculations
in this paper because we are interested in delimiting the
densities where the ionization is fast from those where it
is slow.

In a
real gas, the atoms have a random spacing and the distribution
of spacing affects the amount of ionization. In order to control
for this, we performed calculations for the unphysical situation
where the atoms have a fixed spacing. We performed calculations for
particles on a line, on a square array, and on a cubic grid.
For these cases, any
ionization for separations larger than the maximum ionization for
a pair of atoms will necessarily be due to many-body interactions.
We compare these results to calculations for atoms randomly
distributed in space. We found that the character of ionization
substantially changed when going to a random distribution. In
our calculations, the density fluctuations play a larger role
in ionization than the many-body interactions. However, our
results could not reproduce the experimental results which
suggests there is at least one other important mechanism
for ionization in a dense Rydberg gas.

\section{Numerical Method}\label{SecNM}

Our calculations are purely classical where the electrons obey
Newton's equations and the nuclei are fixed in space. We solve
the coupled first order equations in $\vec{v},\vec{r}$ using
an adaptive step size Runge-Kutta algorithm similar to that
in Ref.~\cite{PTV}. The main change is in how we scale the variables
with all of the velocity components of the $i$-th electron
being scaled by the speed of the $i$-th electron and the position
components being scaled by the distance to the closest nucleus.
For each set of initial conditions, we checked the change in
total energy at the final time. If the energy drifted by more than 0.1\%,
the trajectory was rerun with the same initial conditions but
the error scale decreased by an order of magnitude. The process
was repeated until the energy drift was less than our set value.
We tested that our results were converged with respect to the
setting of our accuracy parameter.

We defined ionization to be when any electron reaches a distance
more than 100 atom spacings from the central position of the
many atom system. We chose 2,000 Rydberg periods for the final time of
the calculation. This is long
enough that most of the trajectories that lead to ionization
will have an electron reach the final distance. But it is
not so long that we waste computer time solving trajectories
that will never lead to ionization.

We used a perfect Coulomb force for the electron-electron
interactions, but a soft-core force for the electron-nuclei
interactions. The potential energy between an electron and
nucleus was proportional to $-1/\sqrt{r^2+b^2}$ where $r$
is the distance between the electron and nucleus and $b$ is
a constant. Because of a non-zero $b$, the potential is not
singular and the force does not diverge as $r\to 0$. Defining
a principal quantum number from the electron's
launch energy using $-13.6\; eV/n^2$, we chose $b= 2 (n/20)^2 a_0$
with $a_0$ the Bohr radius;
this gives a screening length $b$ that is $\sim 1/400$ the size
of the atom. We found that choosing $b=a_0$ gave similar results
at the price of longer calculations.
In all of our calculations, one electron was launched from
near each nucleus. This simulates the photo-excitation step.
For all calculations, we launched the electrons with $n=60$.
Because this classical system scales with $n$ and we present
all our results in terms of ratios, the actual value of $n$
chosen is not relevant. At
small enough $n$, quantum effects will become important but
that is beyond the scope of this paper.

For the grid calculations, the nuclei were exactly on a grid
of points in one, two or three dimensions. For the random
calculations, the $x-$, $y-$, and $z-$positions were chosen
from a flat distribution between 0 and $L$. In Figs.~4-6,
the separation $D$ is defined to be $L$ divided by the number
of atoms for one dimension, $L$ divided by the square root of
the number of atoms for two dimensions, and $L$ divided by the
cube root of the number of atoms for three dimensions.

In an experiment, the electrons are not simultaneously excited
to the Rydberg state but randomly absorb photons proportional to
the time dependent laser intensity. This duration will depend
on the specific laser and the effect will depend on the ratio
of the duration to the classical Rydberg period of the state being
excited. Clearly, we don't want to launch all electrons at the
same time because they will initially have the same phase
in the classical orbit. We did calculations where the time
of each electron's launch was random with a flat distribution
between 0 and 1 Rydberg period or with a flat distribution between
0 and 100 Rydberg periods. We found that the results quantitatively
depended on how we launched the electrons but the qualitative
results we were after did not depend on how the random times
were chosen.

Finally, we chose the initial position of each electron to be
randomly on a sphere of radius $r_0=2 n^2a_0/100$ centered on its
nucleus. The speed of the electron is determined by its
energy $-13.6\; eV/n^2$. The direction of the velocity was chosen
to be perpendicular to the radius making $r_0$ the perigee of
the orbit for an isolated atom. The direction of the velocity
was random in the plane perpendicular to $\hat{r}$ with
$\vec{v}=(\hat{\theta}\cos (\alpha )+\hat{\phi}\sin (\alpha ))v$
where $\alpha$ was from a flat distribution between 0 and $2\pi$.

Of all the choices for initial conditions, our results most
strongly depended on our choice of $r_0$. If instead, we chose a
microcanonical ensemble as in Ref.~\cite{FR1} our ionization
curves shift to smaller atom separation by $\sim 10$\%. As with
other choices described above, the general trends and conclusions
do not depend on $r_0$.

\section{Results}

We performed two styles of calculations to try to cleanly show the
effect of many-body ionization. In one set of calculations,
we have the nuclei on equally spaced points in one dimension,
on a square lattice in two dimensions, or on a cubic lattice
in three dimensions. For these cases, there is a limit on
the atoms' smallest separation and, thus, any increase over
independent pairs is an indication of many-body ionization.
As might be expected, the many-body effect is more apparent
with increasing dimension.

The other set of calculations is to randomly place atoms
on a line, within a square, or within a cube. Besides many-body
ionization, now there can be pairs of atoms that are
randomly close enough to quickly ionize. This second effect
leads to substantially more ionization compared to a grid of
nuclei.

All plots show the probability for ionization as a function
of atom spacing. The probability for ionization is the same as
the fraction of atoms that ionized averaged over all of the calculations
with different initial conditions.

\subsection{Atoms on a Regular Grid}

\begin{figure}
\resizebox{80mm}{!}{\includegraphics{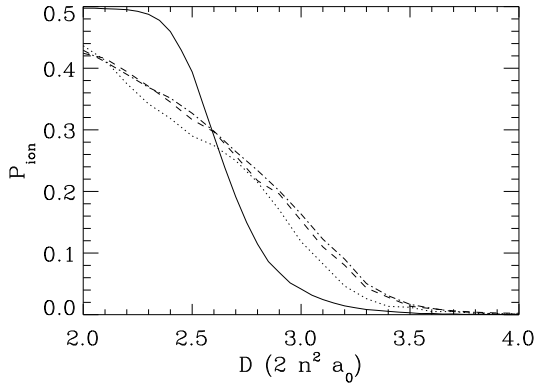}}
\caption{The probability for an atom to be ionized as a function of the
atom separation for atoms in a line. The different line-types
correspond to different number of atoms:
2 atoms (solid), 4 atoms (dotted), 8 atoms (dashed),
and 16 atoms (dash-dot).}
\end{figure}

In Fig.~1, we plot the fraction of atoms ionized
versus the separation of atoms for different number
of atoms. The solid line is for a pair of atoms. We can compare
this result to that reported in Ref.~\cite{FR1} by multiplying
the curve in Fig.~1 by a factor of two because the fraction of
trajectories leading to ionization is just two times the fraction
of atoms ionized for calculations with a pair of atoms. The present
result slightly differs from that reported in Ref.~\cite{FR1}
in that 90\% or more of the trajectories lead to ionization for
scaled separations of 2.3 in Fig.~1 while the value was 2.1
in Ref.~\cite{FR1}. This difference is due to the choice made
for the initial electron conditions as discussed in
Sec.~\ref{SecNM}.

The results for the two atom case have the simplest explanation.
As the separation decreases, the probability that at least one atom
will ionize rises to nearly one. There is a rapid drop in ionization
probability between 2.5 and 3.0 which reflects the decreased coupling
between the atoms and the destruction of the resonance condition
as energy is exchanged between atoms. As one atom gains energy,
its Rydberg period increases while the Rydberg period for the atom
that loses energy decreases. When atoms are widely separated, this
destruction of the resonance condition prevents ionization.

There is a large change in the ionization probability when going from
2 to 4 atoms. For small separation, there is a {\it decrease} in the fraction
of atoms ionized. This is because the atoms might not ionize in ordered
pairs. For example, atoms 2 and 3 might quickly ionize in the 4 atom case
leaving atoms 1 and 4 far away from atoms that they can strongly
interact with. This only needs to occur in approximately 20\%
of the runs to obtain the effect seen in Fig.~1. For larger separation,
there is an {\it increase} in the ionization fraction due to
many-body ionization as discussed in Ref.~\cite{THS}.
Compared to the two atom case, the atoms have
more near neighbors. There is a greater chance for exchanging energy.
Also, the destruction of the resonance condition for the two atom
case does not necessarily hold for more atoms. For example, atom 2
could gain energy from atom 1 while atom 3 gains energy from atom 4;
this will leave atoms 2 and 3 in (near) resonance and they can
continue the exchange of energy until one of them ionizes.

Note that the 8-atom and 16-atom results
are nearly identical. This shows that the one-dimensional case
quickly converges with respect to the number of atoms. There are
only two atoms at the edge of the grid and, thus, the effect of finite
atom number is small.

We also calculated the fraction of configurations that led to {\it at
least one} ionization. This is not directly related to an
experimental observable, but lends itself to an easy test
of many-body ionization.
As an example, the 4 atom case has 3 pairs
of atoms with a separation $R$. The probability for at least one
ionization if
each pair independently ionizes is
one minus the probability for all pairs to {\it not}
ionize. Even for this case, we found that there is more ionization than can
be accounted for simply by the increased number of pairs of
atoms. Thus, there must be some cooperativity in the ionization
process which can be counted as many-body ionization.
However, we found that a large part of
the increase is simply due to the increase in number of atom
pairs.

\begin{figure}
\resizebox{80mm}{!}{\includegraphics{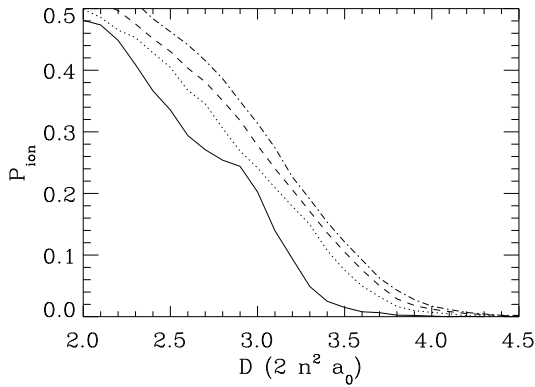}}
\caption{The probability for ionization as a function of the
atom separation for atoms in a square lattice. The different line-types
correspond to different number of atoms:
$2^2$ atoms (solid), $3^2$ atoms (dotted), $4^2$ atoms (dashed),
and $5^2$ atoms (dash-dot).}
\end{figure}

Figure~2 shows the fraction of ionized atoms in a square array
versus their separation for different number of
atoms. Note there is a slightly larger $x$-range
in Fig.~2 compared to Fig.~1. Unlike the one-dimensional case,
the fraction of atoms ionized increases with the number of atoms
for the full range of separations shown. Also, there does not
seem to be convergence with respect to the number of atoms in
a simulation. This is partly due to the larger fraction of atoms
on the surface of the grid. Even the case with $5^2$ atoms has
more surface atoms than interior atoms:
4 corner atoms, 12 edge atoms, and 9 interior atoms.

As expected, there is more ionization for the
two dimensional case compared to one dimension because
each atom has more neighbors that are close. This leads to a net stronger
interaction and, thus, a larger fraction of atoms ionize.
To quantify the increase of ionization, we note that approximately
10\% of the atoms ionize for the 
one dimensional case with 16 atoms at a separation $D=3.2$ compared to
the two dimensional case with 25 atoms at a separation $D=3.6$. Due
to the lack of convergence with respect to atom number, we do not
have a firm prediction of the large atom limit of the fraction of
ionized atoms.

\begin{figure}
\resizebox{80mm}{!}{\includegraphics{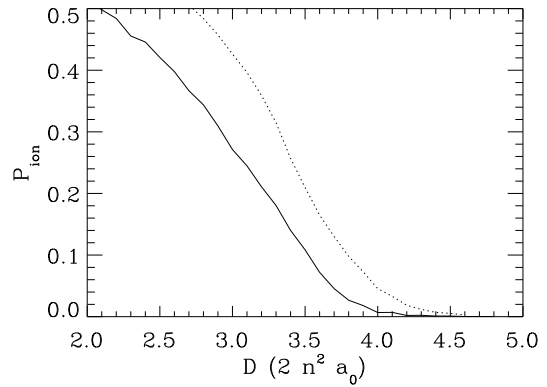}}
\caption{The probability for ionization as a function of the
atom separation for atoms in a cubic array. The different line-types
correspond to different number of atoms:
$2^3$ atoms (solid) and  $3^3$ atoms (dotted),}
\end{figure}

Figure~3 shows the fraction of ionized atoms in a cubic array
versus their separation for different number of
atoms. Note there is a slightly larger $x$-range
in Fig.~3 compared to Figs.~1 and 2. We only have two examples
of a cubic array because the number of atoms increases very
rapidly in three dimensions and the computer time scales approximately
with the third power of the number of atoms. Because there is
such a large change between the $2^3$ and $3^3$ cases, we can
not predict how large an increase in ionization would be present
for a large number of atoms.

Even though the three dimensional case is not converged, it's clear
that there is more ionization than in the one and two dimensional
cases. Approximately 10\% of the atoms ionize for the three
dimensional case with 27 atoms at a separation $D=3.8$. If we
compare to the one dimensional case with 2 atoms ($D=3.0$), the
increase in separation does not appear to be very large (i.e.
approximately 25\%). However, converting to a change in density
by cubing the ratio gives a factor of 2.

One of the signs of many-body ionization is the ionization
that occurs for larger separation. The ionization
fraction for a pair of ions is less than 1\% for separations larger
than $3.5\times 2 n^2 a_0$. However, for more atoms and higher dimensions,
there can be substantial ionization for separations larger than this
value. In fact, the ionization fraction is approximately 20\% for 
the three dimensional case with $3^3$ atoms and this separation.
This highlights the cooperativity that can occur during ionization.

\subsection{Randomly Placed Atoms}

In this section, we present results when the atoms are randomly
placed in a $d-$dimensional region. The position of the ions
are from a flat random distribution between 0 and $L$ in each
dimension.
The separation $D$ is defined to be $D=L/N^{1/d}$ where $N$
is the number of atoms.

\begin{figure}
\resizebox{80mm}{!}{\includegraphics{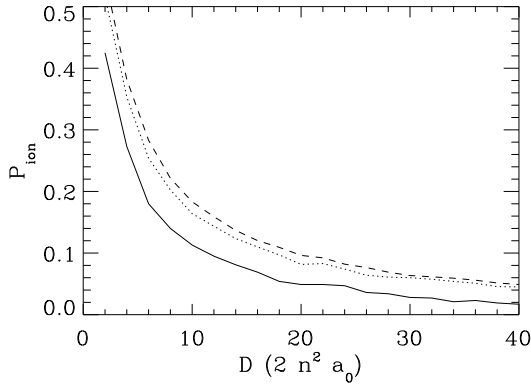}}
\caption{Same as Fig.~1 but for random placement of atoms. Note
the vastly different range for $D$.}
\end{figure}

In Fig.~4, we plot the fraction of atoms ionized
versus the separation of atoms randomly placed on a line.
There is an order of magnitude difference in the range of
separation compared to Fig.~1. This difference reflects the qualitative
change in ionization when the atoms are randomly placed. Even
for large separation, there can randomly be pairs of atoms that
are close enough to ionize. If this interpretation is correct,
the ionized fraction should be proportional to $1/D$ for large separation.
In fact, the simple function $2/(D+1)$ is a good approximation to the
fraction of ionized atoms for $D\geq 5$.

For one dimension, there is an enormous effect from the random placement,
but there is also some effect from the many-body interactions. This is
reflected in the increase of ionization with number of atoms. For example,
there is 10\% ionization at $D\simeq 11$ for 2 atoms, at $D\simeq 17$ for 4 atoms,
and $D\simeq 19$ for 8 atoms.

\begin{figure}
\resizebox{80mm}{!}{\includegraphics{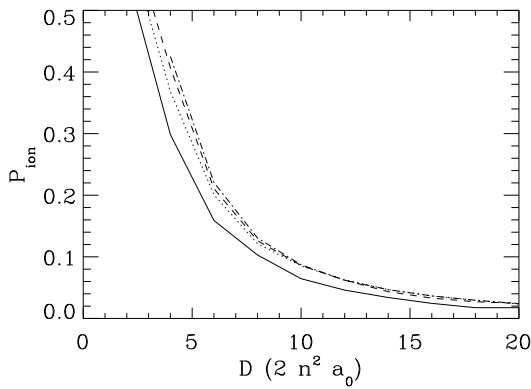}}
\caption{Same as Fig.~2 but for random placement of atoms. Note
the vastly different range for $D$.}
\end{figure}

We plot the fraction of atoms ionized versus the separation of atoms
randomly placed inside a square in Fig.~5. Again, there is a large
increase in the range of $D$ shown compared to Fig.~2. As with
the results in Fig.~4, we can attribute this difference to the random
placement of atoms and the possibility that random pairs of atoms can be
close enough to ionize. If this is the major effect, the ionized fraction
should be proportional to $1/D^2$. We found that the function
$11/(D+1)^2$ is a good approximation to the fraction of atoms ionized
for $D\geq 5$.

\begin{figure}
\resizebox{80mm}{!}{\includegraphics{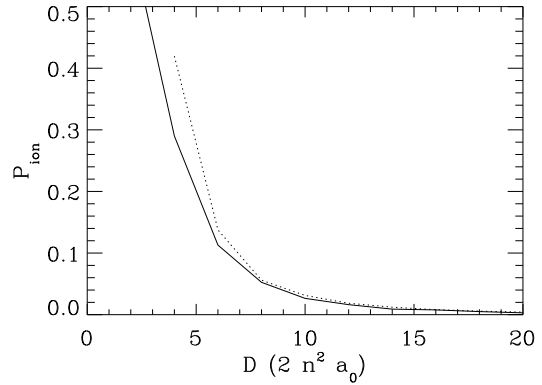}}
\caption{Same as Fig.~3 but for random placement of atoms. Note
the vastly different range for $D$.}
\end{figure}

We plot the fraction of atoms ionized versus the separation of atoms
randomly placed inside a cube in Fig.~6. Again, there is a large
increase in the range of $D$ shown compared to Fig.~2. As with the
one and two dimensional cases, the density fluctuation appears to
be the largest effect. We found that the function
$38/(D+0.6)^3$ is a good approximation to the fraction of atoms
that are ionized. There appears to be some effect from many-body
ionization but it is difficult to discern when the smaller calculation
already has 8 atoms.

To compare the effect of many-body ionization and of fluctuation
on the ionization, we can compare the separation where 10\% and
20\% of the atoms are ionized for a pair of atoms, a cubic grid of
atoms and atoms randomly placed in a cube. This discussion is tentative
because the cubic grid of atoms does not appear to be converged and, thus,
the separation will be underestimated.
For 10\% ionization, two atoms need a separation of 3.0 compared to 
3.8 for the cubic grid and 6.9 for the atoms randomly placed in a
cube. Taking this as the measure, the many-body ionization allows
a density decrease by a factor of $(3.8/3.0)^3\simeq 2.0$ while the
fluctuations allow an additional decrease by a factor of $(6.9/3.8)^3
\simeq 6.0$. For 20\% ionization, two atoms need a separation of 2.8
compared to 3.6 for the cubic grid, and 5.6 for the atoms randomly
placed in a cube. Taking this as the measure, the many-body ionization
allows a density decrease by a factor of $(3.6/2.8)^3\simeq 2.1$ while the
fluctuations allow an additional decrease by a factor of $(5.6/3.6)^3
\simeq 3.8$. By either measure, the random placement has the larger
effect on ionization although the many-body interaction is {\it not}
negligible.

\subsection{Comparison with experiment}

Reference~\cite{THS} found substantial ionization for densities
much smaller than the base density defined in Eq.~(\ref{eqDen}).
We can use the results of our calculation for atoms randomly
placed in a cube as a comparison. The experiment had different
amounts of ionization for somewhat different cases.
We will use the density at 10\% ionization as our benchmark
density; the answer does not qualitatively change if we use
a somewhat higher ionization fraction as the benchmark. We obtain
a density of
\begin{equation}
\rho_{10\%\;ion} = [(256\pi /3)/(2\times 6.9)^3] \rho \simeq 0.1\rho .
\end{equation}
Thus, we obtain substantial ionization for a density an order of
magnitude smaller than the base density whereas Ref.~\cite{THS}
had substantial ionization for densities two orders of magnitude smaller than
the base density. While an absolute number for the density is
hard to obtain experimentally, it seems unlikely to us that the
measurement would be wrong by an order of magnitude.

To decrease the density by an order of magnitude, we would need to
increase the separation from 6.9 to approximately 14. For this
separation we obtain approximately 1\% ionization.
We have considered two possible mechanisms, not in our calculations,
which could increase the fraction of ionized atoms. The first
is electron collisions. The $\sim 1$\% of promptly ionized
electrons could be bound by the space charge effect and then
an ionization cascade as in
Refs.~\cite{RTN,RH1,PPR,LNR,LTG,ARW,MRK,ZFZ,VHS,SOM}
could occur. An argument against this mechanism is that the ionization
cascade is typically a slow process compared to the time scales
in Ref.~\cite{THS}. However, the
Refs.~\cite{RTN,RH1,PPR,LNR,LTG,ARW,MRK,ZFZ,VHS,SOM}
required time to build up the space charge. Perhaps, the prompt
ionization in Ref.~\cite{THS} allows the ionization cascade to
start almost instantly. The second mechanism that might be possible
is the formation of fast atoms and ions during the ionization step. The
calculation of Ref.~\cite{FR1} and the experiment of Ref.~\cite{KR1}
observed that a Penning ionization led to fast Rydberg atoms with a kinetic
energy $\sim 1/5$ of the original binding energy of the cold Rydberg
atoms. These fast Rydberg atoms could collide with the much more
numerous cold Rydberg atoms causing additional Penning ionization
events. A somewhat more complicated variation of this mechanism involves
the fast ion undergoing a charge exchange with a cold Rydberg atom
which leads to a fast Rydberg atom that can collide with other
atoms giving Penning ionization. Performing a realistic simulation
of the ionization cascade or the fast Rydberg collisions is beyond
the scope of this paper.

\section{Conclusions}

We have performed classical calculations of prompt ionization in a
frozen Rydberg gas. Our calculation fixed the position of the ions
but allowed for the full motion of the electrons. The calculations
were inspired by the measurements in Ref.~\cite{THS} which showed
substantial ionization for densities two orders of magnitude smaller
than a reasonable base density. They attributed the increase in
ionization to many-body ionization.

We performed calculations for atoms on a grid and atoms randomly
placed within the same volume. By comparing the two calculations,
we attribute a factor of $\sim 2$ increase to many-body ionization
and a factor of $\sim 5$ increase to fluctuation in nearest neighbor
separations. We can not account for the extra factor of $\sim 10$
observed in Ref.~\cite{THS}, but we briefly discussed two possible
mechanisms that could increase the ionized fraction of atoms.

This
work was supported by the Chemical Sciences, Geosciences, and
Biosciences Division of the Office of Basic Energy Sciences, U.S.
Department of Energy and by the Office of Fusion Energy, U.S.
Department of Energy.

\end{document}